 \documentclass[11pt]{article}

 \pdfoutput=1

\usepackage{graphicx} 
\usepackage{amsmath}

\usepackage{amssymb}

\usepackage{enumitem}

\usepackage[francais,english]{babel}

\usepackage{float}
\usepackage{ifthen}
\usepackage{hyperref}

\begin{document}

\title{{\bf A translation of the paper: \\
On the question of avoiding the infinite ``self-reaction''\footnote{Note of the translator: or ``self-energy.''} \hspace*{-1mm} of the electron,
\\
by V. Ambarzumian and D. Ivanenko (1930).
\vspace*{-8mm} }
}

\author{\noindent\hfil\rule{0.99\textwidth}{.4pt}\hfil \vspace*{1mm} \\
V. Ambarzumian and D. Ivanenko {(1930)}.
\\
Zur Frage nach Vermeidung der unendlichen Selbstrückwirkung des Elektrons.\\
{\it Zeitschrift für Physik,\/} {Vol.64}: p.563--567. \\
(Received 21 July 1930) \\
\url{https://link.springer.com/article/10.1007/BF01397206}\\
\noindent\hfil\rule{0.99\textwidth}{.4pt}\hfil
\vspace*{-2mm} }


\date{Translated by Pascal Marquet / \today. \\
\noindent\hfil\rule{0.99\textwidth}{.4pt}\hfil \vspace*{1mm}}

\maketitle



\vspace*{-12 mm}

\begin{center}

Abstract

An attempt is made to avoid the difficulty of the infinite reaction of the electron on itself, which occurs in quantum electrodynamics, by introducing difference equations instead of differential equations. This vision allows the difficulty of the relativistic wave equation emphasised by Klein,\footnote{O. Klein, ZS. f. Phys., {\bf 45}, 198, 1927.} for example, to be essentially eliminated.
\end{center}
\vspace*{0 mm}

The quantum electrodynamics of Heinsenberg and Pauli\footnote{W. Heinsenberg and W. Pauli, ZS. f. Phys., {\bf 61}, 1, 1929.} leads to an infinite reaction of the electron on itself. The reason for this is the assumption of the point-like electron, which has so far proved so well in quantum mechanics. In classical theory, this difficulty could be overcome by introducing a finite electron radius $r_0$ (although not in a completely flawless way). Such an assumption is not possible in quantum mechanics.

It makes no sense at all to talk about the structure of the electron, since the determination of it must necessarily be based on the measurement of the distances between two points of the electron.
The measurement of a distance on the electron has to be done e.g. by using a ``$\gamma$-ray microscope''. Since the radius of the electron is certainly not larger than $10^{-12}$~cm, we are forced to use beams with a wavelength at least not larger than $10^{-13}$~cm. Then the electron radius is measured as a first approximation. Under the influence of these light quanta, the electron suffers a recoil, i.e. changes its velocity. The magnitude of the change in velocity depends on the direction, but is of the order of the speed of light. Such uncertainty in the velocity for the moment of observation implies a corresponding uncertainty for the calculated length on the electron at rest, since the Lorentz contraction is also indeterminate.

So there is absolutely no point in talking about the shape and structure of the electron in the ordinary sense, because the error in determining a length on the electron is of the order of magnitude of that length itself. It therefore seems necessary to change the whole usual concept of spatial extent for such small particles. Here we would like to propose a very preliminary way out, but which can perhaps serve to seek out a consistent theory that will finally solve the problem of space in quantum mechanics.

\noindent\hfil\rule{0.99\textwidth}{.4pt}\hfil 

(1) let us introduce a cubic integer point lattice in three-dimensional space with the lattice constant $a$, which remains undetermined for the time being.

We require that the electrons can only be located in grid points, i.e. that they should only have the following coordinates:
\begin{equation}
x = k \: a, \hspace*{3mm}
y = m \: a, \hspace*{3mm}
z = n \: a  \hspace*{3mm}
(k, m, n = 0, 1, 2 … -1, -2, ...) 
\nonumber
\end{equation}

We must therefore write down all the equations of atomic mechanics as difference equations. Our difference equations should change into ordinary differential equations of quantum mechanics at the limit $a \rightarrow 0$. Of course, all coordinate differences should remain finite, i.e. our quantum numbers $k, m, n$ should strive towards $\infty$. The solutions of the differential equations must converge to corresponding solutions of differential equations.

Approximately, we can assume that the operator 
$\partial / \partial x$ is replaced by the corresponding difference 
$\varphi_x = 
[ \: \varphi(x+a) \: – \: \varphi(x)  \: ] / a$. 
The form of the equations remains the same.\footnote{R. Courant, K. Friedrichs and H. Loeny, Math. Ann., 100, 32, 1928.}

Our next task is to calculate the interaction of the electron with itself. For this we need to calculate the Green's function 
$g \: (p_1 \: p’)$ of Laplace's equation:
\begin{equation}
\varphi_{xx} \: + \: \varphi_{yy} 
\: + \: \varphi_{zz} \; = \; 0
\end{equation}
at $p = p'$, because the reaction of the electron with itself is the same according to Heisenberg-Paulian quantum electrodynamics: 
\begin{equation}
\varepsilon \; = \; 2 \: e^2 \: g \: (p_1 \: p)
\end{equation}
[\,in (Heisenberg-Pauli) H.P.'s work, only the electrostatic reaction energy is calculated, of the amount $1/2 \: e^2 \: g \: (p_1 \: p)$. But if one also takes into account the magnetic interaction energy, one must take the four times larger value, because with summation according to all states we cannot neglect the magnitudes of the order of magnitude $(v/c)^2$\,].

We are content with an approximate calculation of the value of $g \: (p_1 \: p)$. 
If we assume that at a certain distance the function $g \: (p_1 \: p’)$ has the ordinary character of Coulomb's law, i.e. the form $1/r_{pp}$, we obtain from our difference equation a system of linear equations which defines $g \: (p_1 \: p)$. 
Here, of course, it is assumed that the system is placed in a box that has sufficiently large dimensions. 
We have as many equations as there are points within the area at the edge of which we have prescribed the classical values of $g \: (p_1 \: p')$.

The areas with one, then seven and finally $19$ inner points were taken in sequence and the calculation was carried out. 
The values of $g \: (p_1\: p)$ converged quickly, and the following results are obtained with sufficient accuracy:
\begin{equation}
   g \: (p_1 \: p) \; = \; \frac{3.17}{a} \: .
\end{equation}
From this follows for the self-feedback energy:
\begin{equation}
   \varepsilon \; = \; \frac{ 6.34 \: e^2}{a} \: .
\end{equation}
If we assume that the rest energy of the electron has a purely electromagnetic nature, i.e. 
$m \: c^2 = \varepsilon$, we obtain:
\begin{equation}
  a \; = \; \frac{6.34 \: e^2}{m \: c^2}  \: ,	
\end{equation}
i.e. equal in magnitude to the classical diameter of the electron.

It would of course be of importance to get an explicit expression for the electrostatic interaction law.

\noindent\hfil\rule{0.99\textwidth}{.4pt}\hfil 

(2) From our assumptions follows directly the existence of a maximum packing of space by the electrons. 
For the time being we are forced to assume the same point lattice for the protons as well. Then follows the existence of a maximum possible increase of the potential on a given distance of length $a$. 
If we bring two layers of electrons and protons at such a distance, one must have a certain number of particles in each surface element in order to achieve a prescribed potential jump (say, for example, $2 \: m \: c^2$). 
Let $N$ be the number of particles in the unit area and $l$ the distance between the layers, then the magnitude of the potential jump is given by:
\begin{equation}
  V_2 \: - \: V_1 \; = \; 
  2 \: \pi \: e^2 \: N \: l \: . 
  \nonumber
\end{equation}
If we set $V_2-V_1 \: = \: 2 \: m \: c^2$ and 
$l = a$, we get for this case:
\begin{equation}
  N = \frac{m \: c^2}{\pi \: e \: \varepsilon \: a} \: .
  \nonumber
\end{equation}
But according to (5), 
$(m \: c^2) / ( \pi \: e^2 ) = 6.34 / ( \pi \: a )$, 
and consequently:  
\begin{equation}
  N \; = \; \frac{3.17}{3.14} \;
            \frac{2}{a^2} \: .
\end{equation}

Let us now recall that it was precisely a jump of the size $2 \: m \: c^2$ that had a critical significance in Klein's relativistic treatment of the passage of electrons through a potential wall, namely that in the case of a larger jump the wall was completely transparent.

Now we see that the surface density of the particles required for this is approximately equal to the maximum possible which, according to our ideas, can exist at all (because in each grid point there cannot be more than two particles with opposite spin directions). 
Thus, the introduction of the discontinuous space allows to eliminate Klein's paradox for this case. 

If the distance between two layers is greater than $a$, and thus the maximum potential jump is greater than $2 \: m \: c^2$, the relationships appear quite complicated. 
Because in the time necessary for the passage of the barrier, our potential wall must destroy itself considerably as a result of the play of the electromagnetic forces between particles which generate the potential wall, and as a result of the disintegration of the wave packets of the particles constituting the wall. 
Therefore, the cavity of the potential wall must diminish.

\noindent\hfil\rule{0.99\textwidth}{.4pt}\hfil 

{\it  \underline{Conclusion}\/}. 
The quantisation proposed here is not invariant against arbitrary rotations and shifts of the coordinates, but it seems to us that the possibility of securing the invariance of the equations in a special way is not excluded. Perhaps one must introduce the four-dimensional volume as the simplest quantity and demand that it always contains a certain number of grid points.\footnote{This idea had been expressed by Jordan, who in turn had come to some analogous ideas about the quantisation of space. We are very grateful to Professor P. Jordan for his critical comment on these considerations.}

Another ingenious suggestion to deal with the transformations of the lattice space comes from Dr. H. D. Ursell, who envisaged an analogy with the spin electron. Namely, it can be established that a certain point of the grid after the transformation has a certain probability of being hit at any point of the second grid (so that there is a probability of getting a different integer coordinate). Although translations can be treated strictly in this way, rotations offer greater difficulties.

The question naturally arises as to whether time should also be quantified. 
The answer seems to be necessarily positive. 
Quite apart from the necessary four-dimensional symmetry, the existence of a minimal distance also suggests the assumption of a 
{\it minimal wavelength\/}, 
both for light and for matter; and thus we come to the existence of a 
{\it maximum frequency\/}. 
And the maximum frequency of the light probably means nothing else than the minimum time interval, let's say $\Delta t = (1 / c) \: \Delta x$. 
Let us now try to equate this maximum light frequency to the order of magnitude of the frequency of the light quanta that are produced during the annihilation of protons and electrons. Let us assume
\begin{equation}
\frac{\Delta x}{\Delta t} \; = \; c
\hspace*{5mm}  
\mbox{and} 
\hspace*{5mm}  
\nu_{max} = 
\frac{M \: + \: m}{2 \: h} \: c^2 
\; = \; \frac{1}{\Delta t} 
\: .
\nonumber
\end{equation}
From this it follows: 
\begin{equation}
\frac{M}{m} \; \frac{e^2}{h \: c} \; \approx \; 1
\: .
\nonumber
\end{equation}

The empirical fact of the coinciding magnitude of the fine structure constant and the mass ratio of the protons and electrons can thus be made theoretically understandable. 

We would then like to hope that the considerations sketched here could perhaps contribute to the advancement of the core problems.

\noindent\hfil\rule{0.99\textwidth}{.4pt}\hfil 

{\it \underline{Note on correction from the authors}\/}. Heisenberg attempted an analogue quantization. He succeeded in integrating the difference wave equation for the free electron. This resulted in the very strange property of the maximum eigenvalue. We must sincerely thank Prof. W. Heisenberg for his kind communication.

\noindent\hfil\rule{0.99\textwidth}{.4pt}\hfil 

Kharkov, Physical-Technical Institute.

  \end{document}